\documentclass[a4paper, 12pt, fleqn]{article}
\begin{document}\begin{center}
\textbf{\large EXOTIC MATTER AND SPACE-TIME}
\end{center}
\begin{center}
\textbf{\large $-$New Trends in Cosmo-Particle Physics}
\footnote{An invited talk presented at the International Conference on New Trends in High-Energy Physics, 
Yalta, Crimea(Ukraine), September 10-17, 2005, whose reduced version has been published in the Proceedings, edited by P.N.Bogolyubov,
P.O.Fedosenko,  L.L.Jenkovszky, and Yu.A.Karpenko(Bogolyubov Institute for Theoretical Physics, Kiev, 2005), p.259. This is the extended 
version of the recent contribution to the 12th International Conference \& School \lq\lq Foundation \& Advances in Nonlinear 
Science\rq\rq, Minsk, Belarus, September 27-30, 2004, in the Proceedings, edited by V.I.Kuvshinov and G.G.Krylov(Belarusian State 
University, Minsk, 2005), p. 84.}

\vspace{10mm}
Hidezumi Terazawa
\footnote{E-mail address: \textit{terazawa@post.kek.jp}}

\vspace{5mm}
\textit{Institute of Particle and Nuclear Studies,}

\textit{High Energy Accelerator Research Organization,}

\textit{1-1 Oho, Tsukuba, Ibaraki 305-0801, Japan}

\textit{and}

\textit{Midlands Academy of Business \& Technology (MABT),}

\textit{Melbourne Centre, Melbourne Road, Leicester LE2 0GU, United Kingdom}

\vspace{10mm}

(\textit{Dedicated to Professor Hans A. Bethe, the late founder of nuclear physics})

\vspace{10mm}
\textbf{Abstract}
\end{center}
Exotic forms of matter such as carbon nanofoams, hexalambdas and strange stars, pentaquarks, color-balls, \textit{etc.}
and their relations to current problems in cosmo-particle physics such as dark matter and energy are discussed in some details. 
The contents of this talk include the following:

I. Introduction

II. Exotic Molecules

III. Exotic Nuclei and Strange Stars

IV. Exotic Mesons and Baryons

V. Exotic Quarks, Leptons, and Gauge Bosons

VI. Dark Matter and Energy

VII. Conclusion and Future Prospects.

\vspace{12mm}
\textbf{\large I. Introduction}

\vspace{5mm}
In the twentieth century, the atomism became one of the most important principles in physics: matter consists of molecules, 
a molecule consists of atoms, an atom consists of a nucleus and electrons, a nucleus consists of nucleons, a nucleon consists 
of quarks, and, perhaps, a quark consists of subquarks, the most fundamental constituents of matter[1]. In the twenty-first 
century, it has become one of the ultimate goals in physics to find the substructure of fundamental particles such as quarks, 
leptons, gauge bosons, and Higgs scalars. It is, however, still interesting to find new forms of matter which differ from 
ordinary molecules, atoms, nuclei, hadrons, quarks, and leptons. Recently, such exotic forms of matter as carbon nanofoams[2] and 
pentaquarks[3] have been found experimentally (although some doubdts in the latter evidence). In the near future, more exotic 
forms of matter such as hexalambdas[4] and color-balls[5] would be found. In this talk, I am going to discuss these exotic 
molecules, exotic nuclei, exotic hadrons, and exotic fundamental particles(quarks, leptons, and gauge bosons) in some detail. 

\vspace{5mm}
\textbf{\large II. Exotic Molecules}

\vspace{5mm}
Very recently, a new form of carbon called \lq\lq carbon nanofoam\rq\rq\ has been created at the Australian University in 
Canberra by bombarding a carbon target with a laser capable of firing 10,000 pulses a second. As the carbon reached 
temperatures of around $10,000^{\circ}C$, it formed an intersecting web of carbon tubes, each just a few billionths of 
a meter long. It is a spongy solid that is extremely lightweight and, unusually, attracted to magnets. It is the fifth 
form of carbon known after graphite, diamond and two recently discovered types: hollow spheres, known as buckminsterfullerenes 
or buckyballs, and nanotubes[2].

Suppose that there are $N$ carbon nanotubes whose mass, length, and magnetic moment are $m_i$, $a_i$, and $\mu_i$, respectively 
($i=1,2,3,...,N$). For simplicity, let us assume that each nanotube is taken as a straight string with the ends whose positions 
are $\textbf{x}_i^N$ and $\textbf{x}_i^S$. Then, the motion of nanotubes can be described by the Lagrangian of\[L=\sum_i\frac
{\textbf{p}_i^2}{2m_i}-\sum_{i>j}\frac{\mu_i\mu_j}{a_i a_j}(\frac{1}{|\textbf{x}_i^N-\textbf{x}_j^N|}-\frac{1}{|\textbf{x}_i^N-
\textbf{x}_j^S|}-\frac{1}{|\textbf{x}_i^S-\textbf{x}_j^N|}+\frac{1}{|\textbf{x}_i^S-\textbf{x}_j^S|})\]\[+\sum_i\xi_i
(|\textbf{x}_i^N-\textbf{x}_i^S|-a_i),\]where $\textbf{p}_i$ is the nanotube momentum and $\xi_i$ is a Lagrange multiplier.

Although it seems difficult to discuss a quantum state of large $N$ nanotubes even in the simplest case of equal mass, length, 
and magnetic moment($m_i=m$, $a_i=a$, and $\mu_i=\mu$), we can expect to find a close similarity between a microscopic state of 
carbon nanotubes and a macroscopic state of magnetized iron sand. For example, it is natural to expect that at high temperature 
or after long time a net form of carbon nanofoams encounters a phase transition into a gas form of carbon nanotubes. Not only 
experimental but also theoretical detailed studies of the dynamics of carbon nanofoams are highly desirable since this new form 
of matter may be taken as one of the most useful tools even for medical purpose in the near future[2].

\vspace{5mm}
\textbf{\large III. Exotic Nuclei and Strange Stars}

\vspace{5mm}
A super-hypernucleus is a nucleus which consists of many strange quarks as well as up and down quarks. 
In 1979, I proposed the quark-shell model of nuclei in quantum chromodynamics, presented the effective two-body potential between 
quarks in a nucleus, pointed out violent breakdown of isospin invariance and importance of U-spin invariance in superheavy nuclei, 
and predicted possible creation of \lq\lq super-hypernuclei\rq\rq\ in heavy-ion collisions at high energies, based on the natural
expectation that not only the Fermi energy but also the Coulomb repulsive energy is reduced in such nuclei[4]. A similar idea was 
presented independently and almost simultaneously by Chin and Kerman, who called super-hypenuclei \lq\lq long-lived 
hyperstrange multiquark droplets\rq\rq[6]. Five years later in 1984, the possible creation of such super-hypernuclear matter in bulk 
(on a much larger scale of both mass number and space size) in the early Universe or inside neutron stars was discussed 
in detail in QCD by Witten, who called super-hypernuclear matter \lq\lq quark nugets\rq\rq [7] while the properties of 
super-hypernuclei were investigated in detail in the Fermi gas model by Farhi and Jaffe, who called super-hypernuclei \lq\lq strange 
matter\rq\rq[8]. In a series of papers published in 1989 and 1990, I reported an important part of the results of my investigation 
on the mass spectrum and other properties of super-hypernuclei in the quark-shell model[4].

Let $N_u$, $N_d$, and $N_s$ be the number of $u$\rq s, that of $d$\rq s, and that of $s$\rq s, respectively. Then, 
a nucleus of ($N_u$, $N_d$, $N_s$) has the atomic number, the mass number, and the strangeness given by $Z=(2N_u-N_d-N_s)/3$, 
$A=(N_u+N_d+N_s)/3$, and $S=-N_s$. By noting not only a possible similarity between the effective nuclear potential and the 
effective quark one but also an additional three color-degrees of freedom, I have predicted that the magic numbers in the quark-
shell model are three times the famous magic numbers in the nucleon-shell model ($Z,A-Z=2,8,20,28,50,82,126,...$), \textit{i.e.} 
$N_u,N_d,N_s=6,24,60,84,150,246,378,...$. Therefore, the magic nuclei such as $_2^4He$ and $_8^{16}O$ are doubly magic and superstable 
also in the quark-shell model since $N_u=N_d=6$ for $_2^4He$ and $N_u=N_d=24$ for $_8^{16}O$. What is new in the quark-shell model 
is the expectation that not only certain exotic nuclei with a single magic number such as the \lq\lq dideltas\rq\rq, 
$D\delta^{++++}$ with $N_u=6$ and $D\delta^{--}$ with $N_d=6$, but also certain super-hypernuclei with a triple magic number such as 
the \lq\lq hexalambda\rq\rq, $H\lambda$ with $N_u=N_d=N_s=6$, and the \lq\lq vigintiquattuoralambda\rq\rq, $Vq\lambda$ with $N_u=
N_d=N_s=24$, may appear as quasi-stable nuclei. In fact, in the MIT bag model[9], one can easily estimate the mass of $H\lambda$ 
to be as small as 6.3GeV, which is smaller than $6m_{\Lambda}$($\cong 6.7GeV$). However, in the quark-shell model, there is no 
qualitative reason why the \lq\lq dihyperon\rq\rq\ or \lq\lq H dibaryon\rq\rq, $H$ with $N_u=N_d=N_s=2$, should be quasi-stable or 
even stable. I have made many other predictions including a sudden increase of the $K/\pi$ ratio due to production of 
super-hypernuclei in heavy-ion collisions at high energies[4]. 

In 1990, Saito \textit{et al.} found in cosmic rays two abnormal events with the charge of $Z=14$ and the mass number of $A\cong 370$ 
and concluded that they may be explained by the hypothesis of super-hypernuclei[10]. In order to find whether these cosmic ray events 
are really super-hypernuclei as suggested by the cosmic ray experimentalists, I investigated how the small charge-to-mass-number ratio 
of $Z/A$ is determined when super-hypernuclei are created. In the paper published in 1991, I have shown that such a small charge of 
$3\sim 30$ may be realized as $Z\leq \surd\overline{2A/3}$($\cong 15.7$ for $A=370$) if the super-hypernuclei are created spontaneously 
from bulk super-hypernuclear matter due to the Coulomb attraction[11]. Therefore, the most likely explanation for the abnormal events 
seems to be that they are, at least, good candidates for super-hypernuclei as suggested by Saito \textit{et al.}[10].

However, in the other paper published in 1993, I have suggested the second most likely explanation that they may be \lq\lq technibaryonic 
nuclei\rq\rq\ or \lq\lq technibaryon-nucleus atoms\rq\rq [12]. A technibaryon is a baryon which consists of techniquarks in a bound state 
due to the technicolor force[13]. A technibaryonic nucleus is a nucleus which consists of nucleons and a technibaryon. A technibaryon-
nucleus atom is an atom which consists of a negatively charged technibaryon and an ordinary nucleus in a bound state due to the Coulomb 
force. The technibaryon mass can be expected to be about 2TeV either from scaling of the baryon mass with the color and technicolor 
dimensional parameters $\Lambda_C$ and $\Lambda_{TC}$[14] or from my estimation of the techniquark mass to be about $0.5\sim 0.8TeV$ 
from the PCDC (Partially-Conserved-Dilation-Current) anomaly sum rule for quark and lepton masses[15]. The mass value of about 0.4TeV 
obtained for the abnormal cosmic ray events is much smaller than the expected values for technibaryonic nuclei or technibaryon-nucleus 
atoms. However, this value would not be excluded since large experimental error in determining the masses might be involved. In this 
respect, note that the abnormal cosmic ray event found in 1975 by Price \textit{et al.} may be better explained by a technibaryonic 
nucleus or technibaryon-nucleus atom since their later analysis might indicate the charge of $Z\cong 46$ and the mass number of 
$A\geq 1000$[16]. Also note that the abnormal cosmic ray event found in 1993 by Ichimura \textit{et al.} may be better (but much 
less better) explained by a technibaryonic nucleus or technibaryon-nucleus atom since they reported the charge of $Z\geq 32\pm 2$[17].

More recently, I have proposed the third most likely explanation for the abnormal cosmic ray events that they may be \lq\lq color-balled 
nuclei\rq\rq [18]. A color-ball is a color-singlet bound state of an arbitrary number of gluons[5] or of \lq\lq chroms\rq\rq, $C_{\alpha}$ 
($\alpha =0,1,2,3$), which are the most fundamental constituents of quarks and leptons (called \lq\lq subquarks\rq\rq\ in a generic sense) 
with the color quantum number and which form quarks and leptons together with a weak-isodoublet of subquarks (called \lq\lq wakems\rq\rq), 
$w_i$ ($i=1,2$), in the unified composite model of all fundamental particles and forces[1]. A color-balled nucleus is a nucleus which 
consists of nucleons and a color-ball. The color-ball of ($C_0C_1C_2C_3$) is not only electomagnetically neutral but also weakly neutral. 
However, it strongly interacts with any hadrons due to the van der Waals force induced by the color-singlet state of ($C_1C_2C_3$) as 
baryons, the color-singlet states of three quarks. Its mass may be very large as scaled by the subcolor energy scale $\Lambda_{SC}$ (of 
the order of, say, 1TeV)[19] and its size may be very small as scaled by $1/\Lambda_{SC}$ ($\sim 1/1TeV$) but it may be absolutely stable. 
This extremely exotic particle (which we may call \lq\lq primitive hydrogen\rq\rq) may provide us not only the third most likely 
explanation for the abnormal cosmic ray events but also another candidate for the missing mass in the Universe.

Already in 1971, Bodmer pointed out the possibility that super-hypernuclei (which he called \lq\lq collapsed nuclei\rq\rq) may exist 
on a large scale[20]. He even suggested then that they may explain the missing mass in the Universe. For the last one decade, \lq\lq 
strange stars\rq\rq\ consisting of super-hypernuclear matter have been investigated in great detail not only theoretically but also 
experimentally[20]. If the possible identification of the recently discovered unusual x-ray burster GRO J1744-28 as a strange star by 
Cheng, Dai, Wei, and Lu is right[21], the existence of super-hypernuclear matter or \lq\lq strange matter\rq\rq\ has already been 
discovered by astrophysicists as a gigantic super-hypernucleus or \lq\lq strangelet\rq\rq, the super-hyperstar or \lq\lq strange 
star\rq\rq\ in the Universe, before being discovered by high-energy experimentalists in heavy-ion collisions. I must also mention 
that not only the recent possible identification of the x-ray pulsar Her X-1 as a strange star claimed by Li, Dai, and Wang and of 
the x-ray burster 4U 1820-30 proposed by Bombaci[22] but also the more recent possible identification of the newly discovered millisecond 
x-ray pulsar SAX J1808.4-3658 suggested by Li, Bombaci, Dey, and van den Heuvel[23] seems to be just as reasonable as that of 
GRO J1744-28 by Cheng \textit{et al.}[21]. Very recently, NASA\rq s Chandra X-ray Observatory has found two stars, RXJ185635-3757 which 
is too small (about 11.3km) and 3C58 which is too cold (less than 1 million degrees in Celsius), being most likely strange stars[24].

In September, 1999, just before the BNL RHIC was about to open up a high-energy range of order hundred GeV/nucleon for heavy-ion-heavy-ion 
colliding beams, a rumor shocked the whole world[25]. It said that if a negatively charged stable strangelet were produced by RHIC 
experiments, it would convert ordinary matter into strange matter, eventually destroying the Earth. However, I argued that there is no 
danger of such a \lq\lq disaster\rq\rq\ at RHIC since the most stable configuration of strange matter must have positive electric 
charge thank to the fact that the up quark is lighter than the down and strange quarks[4]. This liberation from such a horrible fear in 
the \lq\lq disaster story\rq\rq\ would remind us of the fact that the very existence of ordinary matter depends on the mass difference 
between the proton and neutron which depends on that between the up and down quarks (which further depends on that between the subquarks, 
$w_1$ and $w_2$)!

Concerning the observation of quark-gluon plasma (QGP) states possibly produced in high-energy heavy-ion collisions, which had been 
claimed strongly by the experimental groups at CERN SPS just before the RHIC was about to operate, no clear indication has yet been 
reported by the experimental groups at BNL RHIC[26]. There have been proposed several signs for productions of QGP states in 
heavy-ion collisions including 1) sudden change of transverse-momentum distribution of produced particles[27], 2) sudden increase of 
the ratios of produced antiparticles ($K^+$,$\overline{p}$, \textit{etc.})[28], 3) sudden suppresion of $J/\psi$ productions, and 4) 
sudden broadening of the widths of produced $\rho\rq s$ and $\omega\rq s$. However, I always argue that all the proposals except for 
the first seem to be ambiguous since they may well indicate something else. In fact, very recently the Japanese nuclear experimental 
group has observed a significant difference in the mass spectra below the omega meson between $p+C$ and $p+Cu$ interactions, 
indicating that the spectral shape of mesons is modified at normal nuclear-matter density[29]. 

Thus, I must conclude this Section by saying that there has not yet appeared a clear indication of exotic nuclei found in high energy 
experiments although there are some candidates reported by the cosmic-ray experimentalists and by the astrophysicists. However, I must 
add that not only the recent discovery of \lq\lq triaxially deformed nuclei\rq\rq [30] but also that of \lq\lq superheavy hydrogens\rq\rq 
[31], $^5H$ and $^7H$, would be something exotic in low-energy nuclear physics.

\vspace{5mm}
\textbf{\large IV. Exotic Mesons and Baryons}

\vspace{5mm}
A pentaquark is a baryon which consists of four quarks and an antiquark. Recently, motivated by the prediction of a baryon with $J^P=
1/2^+$, $Q=+1$, $I=0$, and $S=+1$ at $m=1530MeV$ by Diakonov, Petrov, and Polyakov in the chiral soliton model in 1997, the LEPS and 
DIANA Collaborations have found a narrow resonance with $B=1$, $Q=+1$, $I=0$, and $S=+1$ at $m=1.54\pm 0.01GeV$ and at $m=1539\pm 2MeV$, 
respectively[3], which have been claimed as the first evidence for the pentaquark of $(udud\overline{s})$. Although the evidence for 
this exotic baryon, $\Theta^+$, from experiments of various types such as photoproduction, low-energy kaon-nucleus scattering, 
\textit{etc.}, has been increasing with time, the non-obsevation of $\Theta^+$ in many other experiments has also continued, which 
seems to be a kind of mystery[3]. More recently, the NA49 Collaboration has found other narrow resonances with $B=1$, $Q=-2$ and $0$, and 
$S=-2$ both at $m=1.862\pm 0.002GeV$, $\Xi^{--}(dsds\overline{u})$ and $\Xi^0(dsus\overline{d})$[32], and the H1 Collaboration has 
found yet another narrow resonance with $B=1$, $Q=0$, and the negative charm quantum number at $m=3099\pm 6MeV$, $\Theta_c^0(udud
\overline{c})$[33]. However, it seems also a kind of mystery that these new resonances have not yet been observed by any other experiments[3]. 
As it stands now, I would be rather neutral in accepting their observations of $\Theta^+$ as the first evidence for a new form of matter, 
the pentaquark. 

Theoretically, however, one can expect that there exist not only exotic mesons consisting of two quarks and two antiquarks but also 
exotic baryons consisting of four quarks and an antiquark. It is in principle straightforward (but not so easy) to calculate the masses and 
widths of such exotic hadrons in quantum chromodynamics, the Yang-Mills gauge theory of color SU(3) for strong interactions of
quarks and gluons[34]. For $F$ flavors and three colors of quarks $q_i$ with masses $m_i$ ($i=1,2,3,...,F$), the Lagrangian is given by\[
L_{QCD}=\sum_i\overline{q}_i[i\gamma^{\mu}(\partial_{\mu}-ig\frac{\lambda^a}{2}A_{\mu}^a)-m_i]q_i-\frac{1}{4}(\partial_{\mu}A_{\nu}^a-
\partial_{\nu}A_{\mu}^a+gf^{abc}A_{\mu}^bA_{\nu}^c)^2,\]where $g$ is the strong coupling constant of $SU(3)_c$ gluon fields 
$A_{\mu}^a$ ($a=1,2,3,...,8$), and $\lambda^a$ and $f^{abc}$ ($a,b,c=1,2,3,...,8$) are the Gell-Mann matrices and structure constants of 
SU(3), respectively. In the non-relativistic approximation, the Hamiltonian for a hadron consisting of N quarks and antiquarks with 
the momenta $\textbf{p}_i$ is given by\[H=\sum_i(m_i+\frac{\textbf{p}_i^2}{2m_i})-K_G+\sum_{i>j}V_{ij},\]where $K_G$ is the center-of-mass 
kinetic energy of the hadron and $V_{ij}$ ($i,j=1,2,3,...,N$) is the potential between the i-th and j-th quarks or antiquarks in the 
hadronic matter. In the first approximation of ignoring the interaction between a quark (or an antiquark) and a diquark (or an 
anti-diquark) in the hadronic matter, the potential is approximately given by the one-gluon-exchange potential[35] as\[
V_{ij}=\alpha_s(\frac{\lambda_i^a}{2}\frac{\lambda_j^a}{2})(\frac{1}{r_{ij}}-\frac{\pi}{2}\delta^3(\textbf{r}_{ij})
[\frac{1}{m_i^2}+\frac{1}{m_j^2}+\frac{16(\textbf{s}_i\textbf{s}_j)}{3m_im_j}]\]\[
+\frac{1}{m_im_jr_{ij}^3}[\textbf{s}_i\textbf{s}_j-\frac{3(\textbf{r}_{ij}\textbf{s}_i)(\textbf{r}_{ij}\textbf{s}_j)}{r_{ij}^2}]),\]where 
$\alpha_s$($\equiv g^2/4\pi$) is the strong coupling constant, $\lambda_i^a$ and $\textbf{s}_i$ are the color-SU(3) spin and the spin of the 
i-th quark (or antiquark), respectively, and $r_{ij}$ is the distance between the i-th and j-th quarks (or antiquarks). Even in this 
simlified Hamiltonian model, it is not easy to calculate the masses and widths of pentaquarks without depending on more definite picture on 
the structure of exotic baryons.

There have been proposed two different cluster models of pentaquarks: 1) In the Jaffe-Wilczek model, the four quarks form two scalar and 
color-anti-triplet diquark clusters, which form the color-singlet pentaquark state of $(qq)(qq)\overline{q}$ with the remaining 
color-anti-triplet antiquark[36], and 2) in the Karliner-Lipkin model, there are one scalar and color-anti-triplet diquark cluster and a 
spinor and color-triplet triquark cluster, which form the color-singlet pentaquark state of $(qq)(qq\overline{q})$[37]. In both of these 
models, a relative angular momentum of $L=1$ is assumed between the clusters in order to explain the observed narrow width of the state. 
Therefore, the parity of the pentaquark state is predicted to be positive and the flavor structure is also to be in $8+\overline{10}$ in 
both of these models[38]. 

In order to discuss the mass spectra and decay widths of pentaquarks more quantitatively[3], one further needs additional dynamical machineries 
such as the Skyrme model[39,40], the antisymetrized molecular dynamics (AMD)[41,42], and the QCD sum rule[43,44]. Recently, Jenkins and 
Manohar have given the mass formula for non-exotic and exotic baryons as an expansion in $1/N_c$ (where $N_c$ is the number of colors)[40]. 
More recently, Kanada-En\rq yo, Morimatsu, and Nishikawa have tried to calculate the masses and widths of $\Theta^+(uudd\overline{s})$ in 
a quark model with AMD[42] while Kondo, Morimatsu, and Nishikawa have pointed out that the naive pentaquark correlation functions in the 
extended QCD sum rule include two-hadron-reducible contributions, which have nothing to do with pentaquarks[44]. I would rather conclude this 
Section by emphasizing that much more elaborated studies on possible exotic hadrons may be needed not only experimentally but also theretically 
before one claims the discovery of them. Also, I would like to emphasize that not only experimental studies on possible exotic mesons[45] 
but also theoretical ones in the lattice QCD[46] seem to be more desirable since they may suffer from less ambiguities.

\vspace{5mm}
\textbf{\large V. Exotic Quarks, Leptons, and Gauge Bosons}

\vspace{5mm}
The standard model of fundamental particles and forces is a combination of quantum chromodynamics(QCD) for strong interactions of quarks 
and quantum flavor-dynamics for electroweak interactions of quarks and leptons, the unified $SU(2)_L\times U(1)_Y$ gauge theory of 
Glashow-Salam-Weinberg[47]. For three generations of quarks and leptons, $Q_i\equiv (U_i,D_i)$ and $L_i\equiv (N_i,E_i)$ (where $i=1.2.3$), 
the Lagrangian for the latter dynamics is given by\[
L_{EW}=\sum_i[\overline{Q}_{iL}i\gamma^{\mu}(\partial_{\mu}-ig\frac{\tau^a}{2}A_{\mu}^a-ig\rq \frac{Y_Q}{2}B_{\mu})Q_{iL}\]\[
+\overline{U}_{iR}i\gamma^{\mu}(\partial_{\mu}-ig\rq \frac{Y_U}{2}B_{\mu})U_{iR}
+\overline{D}_{iR}i\gamma^{\mu}(\partial_{\mu}-ig\rq \frac{Y_D}{2}B_{\mu})D_{iR}\]\[
+\overline{L}_{iL}i\gamma^{\mu}(\partial_{\mu}-ig\frac{\tau^a}{2}A_{\mu}^a-ig\rq \frac{Y_L}{2}B_{\mu})L_{iL}\]\[
+\overline{N}_{iR}i\gamma^{\mu}(\partial_{\mu}-ig\rq \frac{Y_N}{2}B_{\mu})N_{iR}
+\overline{E}_{iR}i\gamma^{\mu}(\partial_{\mu}-ig\rq \frac{Y_E}{2}B_{\mu})E_{iR}\]\[
-G_{Ui}(\overline{Q}_{iL}\phi^GU_{iR}+\overline{U}_{iR}\phi^{G+}Q_{iL})-G_{Di}(\overline{Q}_{iL}\phi D_{iR}+\overline{D}_{iR}\phi^+Q_{iL})\]\[
-G_{Ni}(\overline{L}_{iL}\phi^GN_{iR}+\overline{N}_{iR}\phi^{G+}L_{iL})-G_{Ei}(\overline{L}_{iL}\phi E_{iR}+\overline{E}_{iR}\phi^+L_{iL})]\]\[
-\frac{1}{4}(\partial_{\mu}A_{\nu}^a-\partial_{\nu}A_{\mu}^a+g\epsilon ^{abc}A_{\mu}^bA_{\nu}^c)^2
-\frac{1}{4}(\partial_{\mu}B_{\nu}-\partial_{\nu}B_{\mu})^2\]\[
+|(\partial_{\mu}-ig\frac{\tau^a}{2}A_{\mu}^a-ig\rq \frac{B_{\mu}}{2})\phi|^2-\mu^2|\phi|^2-\lambda(|\phi|^2)^2,\]
where $g$ and $g\rq$ are the coupling constants of $S(2)_L$ and $U(1)_Y$ gauge fields $A_{\mu}^a$ (where $a=1,2,3$) and $B_{\mu}$, 
$Y$\rq s are the weak-hypercharges of quarks and leptons,  $G$\rq s are the Yukawa coupling constants of the weak-isodoublet Higgs scalar $\phi$ 
and fundamental fermions (quarks and leptons), and $\mu^2$ and $\lambda$ are the mass parameter and self-coupling constant of $\phi$. 
Since there are not only so many fundamental particles such as the fundamental fermions (quarks and leptons), the gauge bosons (photon, 
weak bosons, and gluons), and the Higgs scalars but also so many arbitrary parameters such as the gauge coupling constants (of $SU(3)_c$, 
$SU(2)_L$, and $U(1)_Y$), the Yukawa coupling constants (of quarks and leptons), and the Higgs scalar mass and coupling parameters, the standard 
model may not be the most fundamental theory but a theory which can be derived as an effective and approximate theory at low energies 
from a more fundamental theory. 

In 1977, we proposed the unified composite model of fundamental particles and forces in which not only quarks and leptons but also gauge bosons and 
Higgs scalars are composites of subquarks, the most fundamental form of matter[1,48]. The minimal supersymmetric composite model of quarks and leptons 
consists of an isodoublet of spinor subquarks with charges $\pm 1/2$, $w_1$ and $w_2$ (called \lq\lq wakems\rq\rq standing for \textbf{w}e\textbf{ak} 
and \textbf{e}lectro\textbf{m}agnetic)[48], and a Pati-Salam color-quartet of scalar subquarks with charges $+1/2$ and $-1/6$, $C_0$ and $C_i$ ($i=
1,2,3$)(called \lq\lq chroms\rq\rq standing for colors)[49]. The spinor and scalar subquarks with the same charge $+1/2$, $w_1$ and $C_0$, may form 
a fundamental multiplet of $N=1$ supersymmetry[50]. Also, all the six subquarks, $w_i$($i=1,2$) and $C_{\alpha}$($\alpha=0,1,2,3$), may have 
\lq\lq subcolors\rq\rq, the additional degrees of freedom, and belong to a fundamental representation of subcolor symmetry[19]. Although the subcolor 
symmetry is unknown, a simplest and most likely candidate for it is $SU(4)$. Therefore, for simplicity, all the subquarks are assumed to be quartet 
in subcolor $SU(4)$. Also, although the the confining force is unknown, a simplest and most likely candidate for it is the one described by quantum 
subchromodynamics (QSCD), the Yang-Mills gauge theory of subcolor $SU(4)$[19]. Note that the subquark charges satisfy not only the Nishijima-
Gell-Mann rule of $Q=I_{w3}+(B-L)/2$ but also the \lq\lq anomaly-free condition\rq\rq\ of $\sum_i Q_{w_i}=\sum_{\alpha} Q_{C_{\alpha}}=0$. The Lagrangian for 
QSCD is simply given by\[
L_{QSCD}=\sum_i\overline{w}_i[i\gamma^{\mu}(\partial_{\mu}-ig\frac{\lambda^a}{2}A_{\mu}^a)-M_i]w_i
+\sum_{\alpha}[|(\partial_{\mu}-ig\frac{\lambda^a}{2}A_{\mu})C_{\alpha}|^2+m_{\alpha}^2|C_{\alpha}|^2]\]\[
-\frac{1}{4}(\partial_{\mu}A_{\nu}^a-\partial_{\nu}A_{\mu}^a+gf^{abc}A_{\mu}^bA_{\nu}^c)^2,\]
where g is the coupling constant of $SU(4)_{sc}$ gauge fields $A_{\mu}^a$ ($a=1,2,3,...,15$), $\lambda^a$ and $f^{abc}$ are the extended Gell-Mann 
matrices and structure constants of $SU(4)$, respectively, and $M$\rq s and $m$\rq s are the wakem and chrom masses.

In the minimal supersymmetric composite model, we expect that there exist at least 36 ($=6\times 6$) composite states of a subquark and 
an antisubquark which are subcolor-singlet. They include 1) 16 ($=4\times 2\times 2$) spinor states corresponding to one generation of quarks and 
leptons, and their antiparticles of\[
\nu = \overline{C}_0w_1,l = \overline{C}_0w_2,u_i = \overline{C}_iw_1,d_i = \overline{C}_iw_2\]
and their hermitian conjugates ($i=1,2,3$), 2) 4 ($=2\times 2$) vector states corresponding to the photon and weak bosons of \[
W^+ = \overline{w}_2w_1;\gamma,Z = \overline{w}_1w_1,\overline{w}_2w_2,\overline{C}_0C_0,\overline{C}_iC_i;W^- = \overline{w}_1w_2\]
or 4 ($=2\times 2$) scalar states corresponding to the Higgs scalars of\[
\phi_{ij}=\left(
          \begin{array}{lr}
           \overline{w}_1w_1 & \overline{w}_2w_1 \\
           \overline{w}_1w_2 & \overline{w}_2w_2
          \end{array}
          \right)\]
($i,j=1,2$) and 3) 16 ($=4\times 4$) vector states corresponding to a) the gluons, \lq\lq leptogluon\rq\rq, and \lq\lq barygluon\rq\rq\ of \[
G^a=\overline{C}_i\frac{\lambda_{ij}^a}{2}C_j;G^0=\overline{C}_0C_0;G^9=\overline{C}_iC_i\]
($i,j=1,2,3$), where $\lambda^a$ ($a=1,2,3,...,8$) is the Gell-Mann matrix of $SU(3)_c$, and b) the \lq\lq vector leptoquarks\rq\rq\ of \[
X_i = \overline{C}_0C_i\]
and the hermitian conjugates ($i=1,2,3$), or 16 ($=4\times 4$) scalar states corresponding to the \lq\lq scalar gluons\rq\rq, \lq\lq scalar 
leptogluon\rq\rq, \lq\lq scalar barygluon\rq\rq, and \lq\lq scalar leptoquarks\rq\rq\ of \[
\Phi_{\alpha\beta} = \overline{C}_{\alpha}C_{\beta}\]
($\alpha,\beta=0,1,2,3$). Quarks and leptons with the same quantum numbers but in different generations can be taken as dynamically different 
composite states of the same constituents. In addition to these \lq\lq meson-like composite states\rq\rq\ of a subquark and an antisubquark, 
there may also exist \lq\lq baryon-like composite states\rq\rq\ of 4 subquarks which are subcolor-singlet. These exotic quarks, leptons, and 
gauge bosons are a new form of quark-lepton matter, which is a subject to discuss later in this Section.

In the unified subquark model of quarks and leptons[48], it is an elementary exercise to derive the Georgi-Glashow relations[51], \[
sin^2\theta_w=\sum (I_3)^2/\sum Q^2=3/8\] and \[
f^2/g^2=\sum (I_3)^2/\sum (\lambda^a/2)^2=1\]
for the weak-mixing angle ($\theta_w$), the gluon and weak-boson coupling constants ($f$ and $g$), the third component of the isospin ($I_3$), 
the charge ($Q$), and the color-spin ($\lambda^a/2$) of subquarks, without depending on the assumption of grand unification of strong and 
electroweak interactions. The experimental value is $sin^2\theta_w(M_Z)=0.23117(16)$[52]. The disagreement between the value of $3/8$ predicted 
in the subquark model and the experimental value might be excused by insisting that the predicted value is viable as the running value renormalized 
\textit{a la} Georgi, Quinn, and Weinberg at extremely high energies (as high as $10^{15}GeV$), given the \lq\lq desert hypothesis\rq\rq[53].

The CKM quark-mixing matrix $V$[54] is given by the expectation value of the subquark current between the up and down composite states as\[
V_{ud}\sim <u|\overline{w}_1w_2|d>, ...\]
[55]. By using the algebra of subquark currents[56], the unitarity of quark-mixing matrix, $VV^+ = V^+V =1$, has been demonstrated although 
the superficial non-unitarity of $V$ as a possible evidence for the substructure of quarks has also been discussed by myself[57]. In the first-
order perturbation of isospin breaking, we have derived the relations of\[
V_{us}=-V_{cd}^*,V_{cb}=-V_{ts}^*, ...,\]
which agree well with the experimental values of $V_{us}=0.221\sim 0.227$ and $V_{cd}=0.221\sim 0.227$[52], and some other relations such as\[
V_{cb}(=V_{ts})\cong (m_s/m_b)V_{us}\cong 0.021,\]
which roughly agree with the recent experimental value of $V_{cb}=0.039\sim 0.044$[52]. In the second-order perturbation, the relations of\[
V_{ub}\cong (m_s/m_c)V_{us}V_{cb}\cong 0.0017\] and \[
V_{td}\cong V_{us}V_{cb}\cong 0.0046\]
have been predicted. The former relation agrees remarkably well with the recent experimental data $V_{ub}\cong 0.0029\sim 0.0045$[52]. The 
prediction for $V_{ts}$ and $V_{td}$ also agree fairly well with the experimental estimates from the assumed unitarity of $V$, $V_{ts}\cong 
0.037\sim 0.043$ and $V_{td}\cong 0.0048\sim 0.014$[52]. To sum up, we have succeeded in predicting all the magnitudes of the CKM matrix 
elements except for a single element, say, $V_{us}$.

For the last two decades, we have been trying to complete an ambitious program for explaining all the quark and lepton masses by 
deriving many sum rules and/or relations among them and by solving a complete set of the sum rules and relations[58]. By taking the first 
generation of quarks and leptons as almost Nambu-Goldstone fermions due to spontaneous breakdown of approximate supersymmetry between 
a wakem and a chrom[59], and the second generation of them as quasi Nambu-Goldstone fermions, the superpartners of the Nambu-Goldstone bosons due to 
spontaneous breakdown of approximate global symmetry[60], we have not only explained the hierarchy of quark and lepton masses,\[
m_e\ll m_{\mu}\ll m_{\tau},m_u\ll m_c\ll m_t,m_d\ll m_s\ll m_b,\]
but also obtained the square-root sum rules for quark and lepton masses[61],\[
m_e^{1/2}=m_d^{1/2}-m_u^{1/2}\] and \[
m_{\mu}^{1/2}-m_e^{1/2}=m_s^{1/2}-m_d^{1/2},\]
and the simple relations among quark and lepton masses[62],\[
m_em_{\tau}^2=m_{\mu}^3\] and \[
m_um_s^3m_t^2=m_dm_c^3m_b^2,\]
all of which are remarkably well satisfied by the experimental values and estimates. By solving a set of these two sum rules and two relations, 
we can obtain the following predictions:\[
\left(
\begin{array}{lcr}
m_e & m_{\mu} & m_{\tau} \\
m_u & m_c     & m_t      \\
m_d & m_s     & m_b   
\end{array}
\right) =\]\[
\left(
\begin{array}{lcr}
0.511MeV_{(input)}              &   105.7MeV_{(input)}         &  1520MeV_{(1776.99{+0.29 \atop -0.26}MeV)}  \\
4.5\pm 1.4MeV_{(input)}         &  1350\pm 50MeV_{(input)}     &   183\pm 78GeV_{(178.0\pm 4.3GeV)}  \\
8.0\pm 1.9MeV_{(7.9\pm 2.4MeV)} & 154\pm 8MeV_{(155\pm 50MeV)} &    5.3\pm 0.1GeV_{(input)}
\end{array}
\right) \]
where the \lq\lq inputs\rq\rq\ and the values indicated in the parentheses denote either the experimental data[52] or the phenomenological 
estimates[63], to which our predicted values should be compared. Furthermore, if we solve a set of these two sum rules and two relations, and 
the other two sum rules obtained in the unified composite model of the Nambu-Jona-Lasinio type[64] for all fundamental particles and forces[48],\[
m_W=(3<m_{q,l}^2>)^{1/2}\] and \[
M_H=2(\sum m_{q,l}^4/\sum m_{q,l}^2)^{1/2}, \]
where $m_{q,l}$\rq s are the quark and lepton masses and $<>$ denotes the average value for all the quarks and leptons, we can predict 
not only the four quarks and/or lepton masses such as $m_d$, $m_s$, $m_t$, and $m_{\tau}$ as above but also the Higgs scalar and weak 
boson masses as \[
m_H\cong 2m_t=366\pm 156GeV\] and \[
m_W\cong \surd\overline{3/8}m_t=112\pm 24GeV,\]
which should be compared to the experimental value of $m_W=80.423\pm 0.039GeV$[52]. Recently, I have been puzzled by the \lq\lq new 
Nambu\rq s empirical quark-mass formula\rq\rq\ of $M=2^nM_0$ with his assignment of $n=0,1,5,8,10,15$ for $u,d,s,c,b,t$[65], which 
makes my relation of $m_um_s^3m_t^2=m_dm_c^3m_b^2$ exactly hold. More recently, I have been more puzzled by the relations of 
$m_um_b\cong m_s^2$ and $m_dm_t\cong m_c^2$ suggested by Davidson, Schwartz, and Wali(D-S-W)[66], which can coexist with my relation and which 
are exactly satisfied by the Nambu\rq s assignment. If we add the D-S-W relations to a set of our two sum rules, our two relations, and our 
sum rule for $m_W$ and if we solve a set of these seven equations by taking the experimental values of $m_e=0.511MeV$, $m_{\mu}=105.7MeV$, 
and $m_W=80.4GeV$ as inputs, we can find the quark and lepton mass matrix of \[
\left(
\begin{array}{lcr}
0.511MeV_{(input)}             &   105.7MeV_{(input)}          &  1520MeV_{(1776.99{+0.29 \atop -0.26}MeV)}  \\
3.8MeV_{(4.5\pm 1.4MeV)}       &   970MeV_{(1350\pm 50MeV)}    & 131.3\pm 0.2GeV_{(178.0\pm 4.3GeV)} \\
7.2MeV_{(8.0\pm 1.9MeV)}       &   150MeV_{(155\pm 50MeV)}     &    5.9GeV_{(5.3\pm 0.1GeV)}
\end{array}
\right)\]
where an agreement between the calculated values and the experimental data or the phenomenological estimates looks reasonable. This result may 
be taken as one of the most elaborated \lq\lq modern developments in elementary particle physics\rq\rq. To sum up, we have succeeded in 
explaining and predicting most of the properties (masses and mixing angles) of quarks and leptons in the unified supersymmetric composite model.

Now I am ready to discuss the main subject in this talk: new forms of quark-leptonic matter such as exotic quarks, leptons, and gauge bosons. 
Among \lq\lq meson-like states\rq\rq\ of subquarks, not only the leptogluon $G^0(=\overline{C}^0C^0)$, barygluon $G^9(=\overline{C}_iC_i)$, and 
vector leptoquarks $X_i(=\overline{C}_0C_i)$, but also the scalar gluons, scalar leptogluon, scalar barygluon, and scalar leptoquarks 
$\Phi_{\alpha\beta}(=\overline{C}_{\alpha}C_{\beta})$ are all exotic since they are not familiar fundamental particles accomodated in the standard 
model. The masses of these exotic gauge bosons may be generated by spontaneous breakdown of the $SU(4)_c$ symmetry. In the unified composite model 
of quark-lepton color interactions proposed by myself[67], the $SU(4)_c$ transmuted from $SU(4)_{sc}$[68] is spontaneously broken down to 
$SU(3)_c$ due to the condensation of $\overline{C}_0C_0$ ($<\overline{C}_0C_0>=(V/\surd\overline{2})^2\neq 0$) and the B-L gluon 
$G^{B-L}[=\overline{C}_i(\lambda^{15}/2)_{ij}C_j]$ becomes $\surd\overline{3/2}$ times heavier than the vector leptoquarks $X_i$ (whose mass is 
given by $fV/2$ where $f$ is the gluon couping constant). Since the presently available lower bound on the $X$ mass is $m_X\geq 290GeV$[52], 
the $G^{B-L}$ mass in this model can be bounded by $m_{G^{B-L}}\geq 355GeV$. Since the gluon coupling constant $f$ is determined by 
$f=(4\pi \alpha_s)^{1/2}\cong 1.2$ (for the renormalization point at $m_Z^2$), the vacuum expectation value $V$ can be bounded as 
$V=m_X/f\geq 0.5TeV$. This lower bound on $V$ seems to be consistent with the expectation that the energy scale of QSCD, $\Lambda_{SC}$, 
must be much higher, say, $\Lambda_{SC}\geq 1TeV$, than that of QCD, $\Lambda_C$ ($\cong 0.2GeV$[52]). In short, the physical G-L gluon should 
appear as a neutral, color-singlet, but strongly interacting vector boson which is $\surd\overline{3/2}$ times heavier than the vector leptoqauarks 
and which couples universally with the B-L current of quarks and leptons. This mixture of the leptogluon and barygluon, if exists, may look like 
the source of every matter as it decays strongly into all matter-antimatter pairs. In any case, we expect that all these exotic gauge bosons 
and Higgs scalars must be much heavier than the ordinary weak bosons ($W^{\pm}$ and $Z$) and Higgs scalars($H$\rq s).

More exotic are \lq\lq baryon-like states\rq\rq\ of subquarks. They are subcolor-singlet states of four subquarks including 1) \lq\lq weak 
balls\rq\rq\ of $(w_iw_jw_kw_l)$, 2) \lq\lq pseudo-quarks or pseudo-leptons\rq\rq\ of $(w_iw_jw_kC_{\alpha})$, 3) \lq\lq pseudo-leptoquarks\rq\rq 
of $(w_iw_jC_{\alpha}C_{\beta})$, 4) \lq\lq quasi-quarks or quasi-leptons\rq\rq\ of $(w_iC_{\alpha}C_{\beta}C_{\gamma})$, and 5) \lq\lq color-balls 
of $(C_{\alpha}C_{\beta}C_{\gamma}C_{\delta})$[5], where $i,j,k,l=1,2$ and $\alpha,\beta,\gamma,\delta=0,1,2,3$. The neutral weak ball of 
$\epsilon_{ABCD}w_1^{\uparrow A}w_1^{\downarrow B}w_2^{\uparrow C}w_2^{\downarrow D}$ (where the subcolor indices $A,B,C,D=1,2,3,4$), if any 
although such composite with all $w\rq s$ in the s-states cannot exist as it violates the Fermi statistics, may be very exotic since it has only 
the weak van der Waals force with other weak balls. Since its mass must be as large as $\Lambda_{SC}(\geq 1TeV)$, it can be a good candidate for the 
WIMP (weakly interacting massive particle) responsible for the missing mass in the Universe. Both the pseudo-quarks or pseudo-leptons and 
quasi-quarks or quasi-leptons of $\epsilon_{ABCD}w_1^{\uparrow A}w_2^{\downarrow B}w_i^{\uparrow \downarrow C}C_{\alpha}^D$ and 
$\epsilon_{ABCD}\epsilon_{\alpha\beta\gamma}w_i^AC_{\alpha}^BC_{\beta}^CC_{\gamma}^D$, if any, may look as if they were heavy quarks or leptons 
having the same properties as ordinary quarks or leptons except for their masses of the order of $\Lambda_{SC}$, say, $1TeV$. They can be taken 
as the mirror states of quarks and leptons. The pseudo-leptoquarks of $\epsilon_{ABCD}w_i^Aw_j^BC_{\alpha}^CC_{\beta}^D$, if any although such 
composites with the identical $C\rq s$ ($\alpha=\beta=0,1,2,3$) at the s-state cannot exist because of the Bose statistics, may look like 
vector leptoquarks having the baryon and/or lepton numbers. They can be taken as another source of the baryon and lepton number asymmetry 
in the Universe.

The most exotic is the scalar color-ball of $\epsilon_{ABCD}\epsilon_{\alpha\beta\gamma\delta}C_{\alpha}^AC_{\beta}^BC_{\gamma}^CC_{\delta}^D$. It is 
not only neutral electromagnetically but also neutral weakly, but it has strong interactions with any hadrons due to the van der Waals force 
of the $(C_1C_2C_3)$ content as a baryon, a color singlet state of three quarks $(q_1q_2q_3)$, has strong interactions with any other hadrons. 
It may be called \lq\lq primitive hydrogen\rq\rq\ since it consists of the \lq\lq leptonic subquark\rq\rq, $C_0$, and \lq\lq primitive 
baryon\rq\rq, $(C_1C_2C_3)$. Since it has both the baryon and lepton numbers, it can be taken another source of the baryon and lepton number 
asymmetry in the Universe. Also, since its mass may be as large as $\Lambda_{SC}(\geq 1TeV)$, since its size may be as small as $1/\Lambda_{sc}
(\leq 1/1TeV)$, and since it may be absolutely stable because of baryon and lepton number conservations, it can be another good candidate 
for a particle resposible for the missing mass in the Universe. 

In 1999, I suggested that the Pomeron whose exchange is dominating high-energy hadron-hadron scatterings is also another type of color-ball, 
the color-singlet complex object consisting of an arbitrary number of gluons[5]. It is an extreme form of glueballs in a generic sense 
in comparison with glueballs in a narrow sense, the color-singlet bound states consisting of two gluons, and \lq\lq odderons\rq\rq, 
the color-singlet bound states consisting of three gluons. It seems very difficult to describe such complex objects in quantum field theory 
and to evaluate their effects on hadron scatterings. In fact, Meng, Rittel, Zhang, and company[69] have adopted a statistical approach 
based on the formulation of complex systems by Bak, Tang, and Wiesenfeld[70] in describing the \lq\lq color-singlet gluon clusters\rq\rq\ 
with \lq\lq self-organized criticality\rq\rq\ and used an optical-geometrical method in examining the space-time properties of such objects. 
Here, however, let me discuss how to describe color-balls in an ordinary particle-theoretical way.

In 1978, Nambu and, independently, Tetsuaki Matsumoto showed that, if a meson can be described by the path-ordered phase factor sandwiched 
between a quark and an antiquark in QCD and if the gluon flux is bunched along the path when quarks are largely seperated, the meson can be 
approximated by the Nambu-Goto massless string with quarks at their end points[71]. It was the first field-theoretical demonstration in which 
the Nambu-Susskind hadronic string is realized in QCD. From their demonstration for the realization of hadronic strings it seems natural to 
imagine that, if the gluon flux is not bunched but diversed along the path between quarks, it can be approximated by a \lq\lq hadronic 
(two-dimensional) membrane\rq\rq\ or better by a \lq\lq hadronic (three-dimensional) bundle\rq\rq. In either case, the color-ball can be 
approximately described by the extended Nambu-Goto action of\[
S_{NG}=-\frac{1}{2\pi\alpha '}\int d^n\sigma[-det(g_{\mu\nu}\partial_{\alpha}X^{\mu}\partial_{\beta}X^{\nu})]^{1/2}F^{-(n-2)/2}\] or by the extended 
Polyakov action of\[
S_P=-\frac{1}{2\pi\alpha '}\int d^n\sigma\surd\overline{-h}(\frac{1}{2}h^{\alpha\beta}g_{\mu\nu}\partial_{\alpha}X^{\mu}\partial_{\beta}X^{\nu}
-\frac{n-2}{2}F),\]
where $\sigma_{\alpha}$($\alpha=0,1,...,n-1$) and $X^{\mu}$($\mu=0,1,2,3$) are the world-sheet parameters and the space-time coordinates of 
the hadronic membrane for $n=3$ or the hadronic bundle for $n=4$, respectively, $g_{\mu\nu}=diag(-1,1,1,1)$, $h=det h_{\alpha\beta}$, and $\alpha '$ 
and $F$ are constants. To proceed further for evaluating the effects of color-balls on hadron scatterings in this model seems very difficult 
since it would inevitably make us be involved in some mathematical complexity. However, in this picture of the Pomeron as a color-ball 
we may be able to understand the following properties of the Pomeron found in hadron scatterings at high energies without any manipulations: 
1) The Pomeron cannot be taken as a simple Regge pole but a hadronic membrane or bundle, which is an extension of the Nambu-Susskind hadronic string, 
or a linear combination of an infinite number of Regge poles and Regge cuts, which revives the Freund-Harari hadron duality. 2) The slope of 
the Pomeron trajectory $\alpha_P'$ is much smaller than that of a ordinary Regge trajectory $\alpha_R'$, \textit{i.e.} $\alpha_P'\sim (1/4)\alpha_R'$, 
since the \lq\lq membrane or bundle tension\rq\rq\ is much larger than the string tension. 3) The Pomeron couplings are universal to any flavors of 
quarks (and antiquarks) and, therefore, their factorizability holds since color-balls are not only color-blind but also flavor-insensitive. 
The experimental fact that the Pomeron intercept is larger than unity, \textit{i.e.} $\alpha_P(0)>1$, seems to be very dynamical and cannot be 
explained intuitively even in this picture of the Pomeron. I wish to emphasize the importance of not only further theoretical works but also 
further experimental investigations on the real picture of the Pomeron by checking various things including the ever increasing total, elastic, 
and diffractive cross sections, the universality and factorizability such as the now classic relations of $\sigma(\overline{p}p)\cong \sigma(pp)$, 
$\sigma(\pi p)\cong (2/3)\sigma(pp)$, and $\sigma(\gamma \gamma \to hadrons)\cong \sigma(\gamma p \to hadrons)^2/\sigma(pp)$[72] and the triple-
Pomeron and Pomeron-Reggeon-Reggeon vertices based on the Brandt-Preparata and Mueller diagrams in hadron-hadron, photon (gauge-boson)-hadron, and 
(hadronic) photon (gauge-boson)-photon (gauge-boson) scatterings at high energies[73]. After all, they will eventually clarify one of the most 
fundamental problems in high energy physics, the origin of finite and non-vanishing hadron sizes. In concluding this Section, I also wish to emphasize 
the similarity of the color-ball consisting of chroms and that of gluons in which they are not only electrically neutral but also weakly neutral but 
strongly interacting with any hadrons. They differ from one the other at the points that the former may be stable thanks to baryon and lepton number 
conservation while the latter strongly decays into hadrons and that the former must be much heavier (whose mass $\sim \Lambda_{SC}\geq 1TeV$) and 
much smaller (whose size $\sim 1/\Lambda_{SC}\leq 1/1TeV$) than the latter (whose mass $\sim \Lambda_C\cong 0.2GeV$ and whose size 
$\sim 1/\Lambda_C\cong 10^{-13}cm$).

\vspace{5mm}
In this and the previous Sections, I have discussed exotic forms of matter including an exotic form of atomic matter such as carbon nanofoams, 
the one of nuclear matter such as super-hypernulei and strange stars, the one of hadronic matter such as pentaquarks, 
and the one of quark-leptonic matter such as color-balls. I have introduced the Lagrangian models for these exotic molecules, 
exotic nuclei, exotic hadrons, and exotic quarks, leptons, and gauge bosons in quantum mechanics, in non-relativistic 
quantum mechanics, in relativistic quantum field theory and in general relativistic quantum \lq\lq bundle\rq\rq theory, 
respectively. In order to find all the properties of these exotic forms theoretically, they all have similar complexity, 
the few-body problem, which seems to require much more elaborated studies in the future. Practically, carbon nanofoams seem 
to have very promising future as rosy as carbon nanotubes. As for exotic nuclear matter, the possible direct production of 
super-hypernuclei such as hexalambdas is highly desirable in the future RHIC experiments while the more detailed observation 
of strange star candidates can be expected in the near future. Also, not only experimental searches for exotic baryons 
such as pentaquarks but also those for exotic mesons[45,74] should be extensively continued so that the \lq\lq first 
observation\rq\rq\ of $\Theta^+$ may be confirmed. In concluding this Section, I wish to emphasize that an experimental search 
for color-balls of gluons is as important as that for Higgs scalars since the former are the origin of hadron sizes 
while the latter are the origin of quark, lepton, and gauge-boson masses and that an experimental search for color-balls 
of chroms is even more important since they are the origin of matter!

\vspace{5mm}
\textbf{\large VI. Dark Matter and Energy}

\vspace{5mm}
In this Section, I am going to discuss the relations of these exotic forms of matter to current problems in cosmo-particle physics 
such as dark matter and energy. Before discussing exotic matter as a cadidate for dark matter in the Universe, let me briefly discuss 
ordinary matter such as neutrinos, neutralinos, and white dwarfs.

In 1998, the Super-Kamiokande Collaboration[75] reported the discovery of neutrino oscillation, which indicates not only the violation 
of lepton-number conservation[76] but also the non-vanishing mass of at least one of neutrinos. Recently, it has been almost confirmed 
not only by the K2K Collaboration[77] but also by the SNO Collaboration[78]. Also, in 2000, the DONUT Collaboration[79] established 
the existence of the tau neutrino. Combining these great discoveries together with the other experimental observations such as 
solar and atmospheric neutrino deficits[80], we expect that the electron, muon, and tau neutrino masses are all small and less than 
a few eV even if they are non-vanishing[81]. Therefore, the non-vanishing masses of neutrinos cannot account for the \lq\lq missing 
mass\rq\rq\ in the Universe unless there were another heavier neutrino. 

In 1999, the DAMA Collaboration[82] reported the possible discovery of a weekly interacting  massive object (WIMP) 
although it has not been confirmed by the other experimental groups. Also, in 2001, the Muon (g-2) Collaboration[83] reported 
the possible discrepancy between their new experimental value of the muon g-2 and the current theoretical one from the standard model. 
Although both of these once-claimed anomalies had often been taken as first indications of the super-particles predicted in 
supersymmetric theories[50], it seemed to me too early to decide what would cause them since it could be anything even if they had been real. 
Therefore, it is much too early to claim that neutralinos can account for the \lq\lq missing mass\rq\rq\ in the Universe.

On the other hand, many astronomical observations[84] have recently indicated that the previously undetected white dwarfs may make up 
a substantial fraction of the \lq\lq dark matter\rq\rq\ in the Universe. It may not be too early to accept the most natural expectation that 
the \lq\lq missing mass\rq\rq\ in the Universe mostly consists of the ordinary matter of baryons, in the forms of white (and brown) dwarfs 
and neutron stars[85], in the exotic forms of strange stars[4], \lq\lq technibaryonic nuclear stars\rq\rq[12], and \lq\lq color-balled nuclear 
stars\rq\rq[5,18], or in the form of black holes[86].  In fact, a recent study of the galaxy cluster Abell 2029 using the Chandra Observatory 
agrees with the predictions of cold dark matter and is contrary to other dark matter models[87]. Recently, Paczy\'{n}ski and Haensel have proposed 
that gamma-ray bursts, whose origins have foxed astronomers for decades, might be the signatures of elusive \lq\lq quark stars\rq\rq[88], 
the exotic form of ultra-dense neutron stars and strange stars. More recently, Minchin \textit{et al.}[89] have suggested that VIRGOHI21, the HI source 
detected in the Virgo Cluster survey of Daies \textit{et al.}[90], is a dark halo which does not contain the expected bright galaxy 
and that further sensitive surveys might turn up a significant number of the dark matter halos predicted by dark matter models. Thus, the \lq\lq 
dark matter\rq\rq\ would not be a serious problem compared to the \lq\lq dark energy\rq\rq, \textit{i.e.} the cosmological constant.

Concerning the \lq\lq dark energy\rq\rq\, whose existense was discovered by the Supernova Cosmology Project and the High-Z supernova Search Team 
in 1998[91], I proposed a possible explanation for it in 1999[92]. I have suggested that the \lq\lq cosmological constant\rq\rq\ is not a fundamental 
constant but an effective quantity as induced dynamically, for example, by quantum fluctuations of matter as in pregeometric theories of gravitation[93]. 
In fact, over two decades ago, we showed that not only the Newtonian gravitational constant but also the Einstein cosmological constant 
as well as all the gauge and Higgs coupling constants in the standard model of quarks and leptons can be taken as effective quantities 
induced by quantum fluctuations of matter[94] in the unified composite model (or \lq\lq pregauge and pregeometric theory\rq\rq) of all 
fundamental particles and forces including gravity[95]. If this is the case, neither $G$ nor $\Lambda$ may be a constant at all, 
but both of them may vary depending on the situation or environment as functions of space-time, temperature, presure, \textit{etc.}. 
In fact, over six decades ago, Dirac[96] pointed out such an astonishing possibility of the time dependent Newtonian gravitational \lq\lq constant\rq\rq 
and, recently, many authors including Vilenkin and Starobinsky[97] have discussed such possibility of the variable cosmological term in order to 
solve the dark energy problem. If the cosmological \lq\lq constant\rq\rq\ is not a constant but a variable, there is a no chance for it to always vanish. 
Therefore, it seems very natural to suppose that the cosmological constant had happened to be non-vanishing at the beginning of our Universe and that 
it has never vanished. I have also propose a simple model of our Universe. Define the critical mass-energy density $\rho _c$, \lq\lq 
pressureless-matter-density\rq\rq\ $\Omega _M$, \lq\lq scaled cosmological constant\rq\rq\ $\Omega _{\Lambda}$, and \lq\lq acceleration parameter\rq\rq\ 
$A$ by $\rho _c = 3H^2/8\pi G$, $\Omega _M = \rho /\rho _c = 8\pi G\rho /3H^2$, $\Omega _{\Lambda} = \Lambda /3H^2$, and $A = ad^2 a/dt^2/(da/dt)^2$, 
where $a$ is the cosmological scale parameter in the Friedmann model of the Universe based on the \lq\lq cosmological priciple\rq\rq\ of large-scale 
homogeneity and isotropy consisting of the Friedmann-Robertson-Walker metric of\[
ds^2 = dt^2 - a(t)^2[\frac{dr^2}{1-\kappa r^2} + r^2(d\theta ^2 + sin^2\theta d\phi ^2)],\]
with $\kappa = +1, 0, or -1$ for the closed, flat, or open Universe. Then, the simple model of our Universe is given by $\kappa = 0$ and 
$(\Omega _M, \Omega _{\Lambda}, A) = (0, 1, 1)$ for the early inflationary era, $(1/3, 2/3, 1/3)$ for the radiation dominated era, or 
$(1/3, 2/3, 1/2)$ for the matter-dominated era. This model, which seems to simulate our Universe, suggests that there must be another \lq\lq 
phase transition\rq\rq\ in which $\Omega _{\Lambda}$ changed from 1 to 2/3 in between the early inflationary era and the radiation-dominated era.

Among the recent remarkable cosmological observations, I have been first impressed by the recent observation of the \lq\lq farthest supernova ever seen\rq\rq\ 
by Hubble Space Telescope[98]. \lq\lq This supernova shows us the universe is behaving like a driver who slows down approaching a red stoplight 
and then hits the accelerator when the light turns green.\rq\rq\ Secondly, note that our model of the Universe is also consistent with 
the recent measurement of the cosmological mass density from clustering in the Two-Degree-Field Galaxy Redshift Survey[99] which strongly 
favors a low density Universe with $\Omega_M \cong 0.3$. Thirdly, the CBI Collaboration[100] has found $\Omega_{\Lambda} = 0.64^{+0.11}_{-0.14}$ 
and $\Omega_M + \Omega_{\Lambda} = 0.99\pm 0.12$. Finally, the Wilkinson Microwave Anisotropy Probe(WMAP) team[101] has found: 1) the  first 
generation of stars to shine in the Universe first ignited only 200 million years after the Big Bang, 2) the age of the Universe is 
$13.7\pm 0.2$ billion years old, and 3) $\Omega_M = 0.27\pm 0.04$ and $\Omega_{\Lambda} = 0.73\pm 0.04$. Note that all these observations are 
well described by the simple model of our Universe.

Recently, from the observation of $\gamma$ rays from the Crab nebula, Jacobson \textit{et al.}[102] have essentially ruled out the possible 
violation of Lorentz invariance at Planck energies or below, at least under certain dynamical assumptions in quantum gravity[103]. 
Almost simultaneously, from laboratory experiments, Meyers and Pospelov[104] have placed constraints on Lorentz violation that are not quite 
as strong but are even beyond the Planck scale! Therefore, let me conclude this Section by adding that cosmo-particle physics now seems 
to be evolving into pregeometry[105] or even into prephysics[106] in which such sacred principles in physics as quantum and relativity ones 
are to be questioned or to be explained from more fundamental principles.

\vspace{5mm}
\textbf{\large VII. Conclusion and Future Prospects}

\vspace{5mm}
In this talk, I have discussed exotic forms of matter such as exotic molecules, exotic nuclei and strange stars, exotic mesons and baryons, 
exotic quarks, leptons, and gauge bosons, \textit{etc.}, and their relations to current problems in cosmo-particle physics such as dark matter 
and energy. In this concluding Section, let me discuss new trends and future prospects in high energy physics and cosmology
(or cosmo-particle physics). 

As for exotic nuclei, as I duscussed in Section III, super-hypernuclei or strangelets[4] seem to have been found in the cosmic-ray experiments[10,16,17] 
and candidates for strange stars have been found abundantly[21-24]. Futhermore, one can expect that future RHIC and LHC experiments will be able to 
produce such exotic nuclei and that future astronomical observations will find much more candidates for such exotic stars.

As for exotic hadrons, \lq\lq the elusive pentaquark may be about to disappear\rq\rq\ and \lq\lq mystery deepens as pentaquarks refuse 
to make an appearance\rq\rq[3]. In comparison, as I emphasized in Sections IV and V, exotic mesons have appeared abundantly and  more will appear 
in the near future, serving us for better understanding of quark and gluon dynamics[45,74]. However, it seems more urgent to study the exotic form of 
quarks and gluons called \lq\lq color-ball matter\rq\rq\, \lq\lq quark-gluon liquid\rq\rq, or \lq\lq liquidlike quark-gluon plasma\rq\rq[5] 
both experimentally and theoretically as the recent RHIC experiments discovered a new state of matter not seen since the first moments 
after the big bang[107]. \lq\lq After years of running, RHIC scientists certainly saw something new[108]. Jets of particles flying away 
from the collision seemed to be moving through a group of hard protons and neutrons as would ordinarily be the case. Futhermore, 
after the collision, the system behaved like an expanding puddle of fluid rather than a swarm of particles\rq\rq. \lq\lq It's a ideal 
liquid ... with essentially no viscosity\rq\rq. \lq\lq It is as perfect a fluid as calculations would allow\rq\rq. Therfore, it seems that future 
prospects of high-energy nuclear, hadron, and particle physics may be concentrated into a deep understanding of a single object of color-ball, 
the origin of strong interactions!

As for exotic quarks, leptons, and gauge bosons, as I discussed in Sections III, IV, and V, color-balls[5] may have been already found in the 
anomalous cosmic-ray events[5,10,16.17] or appear always as Pomerons in high-energy particle collisions. In addition, one can expect that future 
HERA, Tevatron, RHIC, LHC, and ILC(or TESLA) experiments will be able to produce not only naked color-balls but also exotic or exited quarks, 
leptons, gauge bosons(including leptogluon, barygluon, and leptoquarks), and Higgs scalars, any one of which will indicate a long-sought 
direct evidence for the substructure of quarks, leptons, gauge bosons, and Higgs scalars. Such an exciting discovery in the future would necessarily 
make us convinced that the unified composite model of all fundamental particles and forces[1] be the final theory in physics!

As for dark matter, as I emphasized in Section VI, there are so many candidates for it including white(and brown) dwarfs, \lq\lq quark stars\rq\rq[88] 
(including ultra-dense neutron stars, strange stars[4], technibaryonic nuclear stars[12], and color-balled nuclear stars[5,18]), and black holes[86] 
that one may not need to assume any WIMPs(weakly interacting massive particles) such as massive neutrinos or massive neutralinos[87]. 
This will be the case if future experimental searches for the WIMPs continue to fail to find any candidates. It will, then, not be a problem at all. 
In fact, the recently observed dark halo, VIRGOHI21, does not seem to consist of anything exotic[89]. Neither will dark energy be a problem, 
as I discussed in Section VI, if it is induced by the quantum fluctuations of matter in the vacuum as in pregeometric theories of gravitation[93,105] 
and if it is not a constant but may vary as the other fundamental physical constants such as the fine-structure and gravitational constants 
in pregaugeometric theories of all fundamental forces[94,95,105]!

Recently, I have proposed a theory of special inconstancy[109] in which some fundamental physical constants such as the fine-structure and gravitational 
constants may vary in pregaugeometry[105]. In the theory, the alpha-G relation of \[\alpha=3\pi/[16\ln(4\pi/5GM_W^2)]\] between the varying fine-structure 
and gravitational constants (where $M_W$ is the charged weak boson mass) is derived from the hypothesis that both of these constants are related 
to the same fundamental length scale in nature. Furthermore, it leads to the prediction of $\dot{\alpha}/\alpha=(-0.8\pm2.5)\times10^{-14}yr^{-1}$ 
from the most precise limit of $\dot{G}/G=(-0.6\pm2.0)\times10^{-12}yr^{-1}$ by Thorsett[110], which is not only consistent with the recent observation 
of $\dot{\alpha}/\alpha=(0.5\pm0.5)\times10^{-14}yr^{-1}$ by Webb \textit{et al.}[111] but also feasible for future experimental tests. In special 
inconstancy, I have also explained the past and present of the Universe and predicted the future of it, which is quite different from that in the 
Einstein theory of gravitation. 

In the principle of special inconstancy, we do not assert that physical constants may vary as a function of time but do that they may vary in general, 
depending on any parameters including the cosmological time, temperature, \textit{etc.}. What is the origin of varying the physical constants? 
The answer to this question may be related to the answer to another fundamental question: 
What is the origin of the fundamental length scale $\Lambda^{-1}$ in nature? It can be spontaneous breakdown of scale-invariance 
in the Universe, which has been proposed by myself[112] for the last quarter century. It can be the natural, dynamical, automatic, \textit{a priori}, 
but somewhat \lq\lq wishful-thinking\rq\rq\ cut-off at around the Planck length $G^{1/2}$ where gravity would become as strong as electromagnetism, 
which was suggested by Landau in 1955[113]. It can also be due to the Kaluza-Klein extra dimension[114], which is supposed to be compactified 
at an extremely small length scale of the order of $G^{1/2}$ or at a relatively large length scale of the order of $1/TeV$ recently emphasized by 
Arkani-Hamed \textit{et al.}[115]. It seems, however, the most natural and likely that the origin of the fundamental length comes from the substructure of 
fundamental particles including quarks, leptons, gauge bosons, Higgs scalars, \textit{etc.}[1]. In the unified composite model of all fundamental 
particles and forces[1], the fundamental energy scale $\Lambda$ in pregaugeometry can be related to some even more fundamental parameters such as 
the masses of subquarks, the more fundamental constituents of quarks and leptons, and the energy scale in quantum subchromodynamics, the more fundamental 
dynamics confining subquarks into a quark or a lepton. In either way, the fundamental length scale $\Lambda^{-1}$ can be identified with the size of 
quarks and leptons, the fundamental particles. 

The space-time with such natural cut-off of $\Lambda^{-1}$ would look exotic not only at short distances of the order of the Planck length $G^{1/2}$ 
but also at long distances of the order of the Hubble parameter inverse $H^{-1}$. In fact, very recently, Aurich \textit{et al.}[116] have pointed out 
that the WMAP data[101] is consistent with the Picard model for a closed Universe[117]. Having too much freedom with varying fundamental physical 
constants and having too little knowledge on the past and present of the Universe, we can predict anything we want or we cannot predict anything definite 
about the future of the Universe! All I can foresee is that cosmo-particle physics may be evolved through cosmo-particle chemistry into cosmo-particle 
biology since we, human-beings, seem to be most interested in the following three questions: \lq\lq Where do we come from?\rq\rq, \lq\lq Where are we 
now?\rq\rq, \lq\lq Where are we going (to)?\rq\rq.

\vspace{45mm}
\textbf{\large Acknowledgements}

\vspace{5mm}
The author would like to thank Professor L.L.Jenkovszky and the other organizers for inviting him to this 
International Conference on New Trends in High-Energy Physics, Yalta, Crimea(Ukraine), September 10-17, 2005, and for their warm 
hospitality extended to him during his stay in Crimea. This talk has been dedicated to Professor Hans A. Bethe, the late founder of 
nuclear physics and astronomy, from whom he had learned not only how to live in this wonderful physical society in the world but also 
how to survive in this awful nuclear age on the earth and in the space during his stay as instructor-research associate in Laboratory 
of Nuclear Studies, Cornell University from 1969 to 1971.

\end{document}